\documentstyle[11pt,epsf,a4]{article}

\newcommand{\Ikdv}{\int\,\frac{{\rm d^4}k}{(2\pi)^4}\;}

\newcommand{\Tr}{\makebox{ Tr }}

\newcommand{\GeV}{\makebox{ GeV}}

\newcommand{\fm}{\makebox{ fm}}

\newcommand{\beq}{\begin{equation}}

\newcommand{\enq}{\end{equation}}

\newcommand{\beqa}{\begin{eqnarray}}

\newcommand{\enqa}{\end{eqnarray}}

\newcommand{\nn}{\nonumber}

\newcommand{\labelm}[1]{\label{#1}}

\newcommand{\lbfi}[1]{\labelm{#1}\end{figure}}

\newcommand{\lbq}[1]{\labelm{#1}\enq}

\newcommand{\lbqa}[1]{\labelm{#1}\enqa}

\newcommand{\befi}[1]{\begin{figure}[ht] \leavevmode \centering
\epsffile{#1.eps}}

\newcommand{\eq}[1]{eq.(\ref{#1})}

\newcommand{\fig}[1]{fig.(\ref{#1})}

\newcommand{\lbcap}[3]{\begin{minipage}{#1}\caption{\small #2}
\labelm{#3}\end{minipage}\end{figure}}

\newcommand{\lbtab}[3]{\centering\begin{minipage}{#1}
\caption{\small #2}\labelm{#3}\end{minipage}\end{table}}

\newcommand{\pa}{\partial}

\newcommand{\cF}{\mbox{$\cal F$}}

\newcommand{\cP}{\mbox{$\cal P$}}

\newcommand{\bA}{\mbox{\bf A}}

\newcommand{\bW}{\mbox{\bf W}}

\newcommand{\al}{\alpha}

\newcommand{\ga}{\gamma}

\newcommand{\de}{\delta}

\newcommand{\ka}{\kappa}

\newcommand{\la}{\lambda}

\newcommand{\rh}{\rho}

\newcommand{\si}{\sigma}

\newcommand{\ch}{\chi}

\newcommand{\om}{\omega}

\newcommand{\De}{\Delta}

\newcommand{\Ph}{\Phi}

\newcommand{\Ps}{\Psi}

\begin{document}
\title{\vspace*{-2cm}\begin{flushright}\normalsize HD-THEP-97-34
\end{flushright}
\vspace*{+2cm}\LARGE \bf Diffractive color-dipole nucleon scattering}
\author{
Michael Rueter\thanks{supported by the Deutsche Forschungsgemeinschaft}
, H.G.~Dosch\\[.7cm]
\it Institut f\"ur Theoretische Physik\\
\it Universit\"at Heidelberg\\
\it Philosophenweg 16, D-69120 Heidelberg, FRG\\[.3cm]
\it e-mail: M.Rueter@ThPhys.Uni-Heidelberg.DE\\
\it \hspace*{1.4cm}H.G.Dosch@ThPhys.Uni-Heidelberg.DE
}
\date{}
\maketitle
\thispagestyle{empty}
\begin{abstract}
We determine the diffractive scattering amplitude of a color-dipole
on a nucleon using a non-perturbative model of QCD which contains only
parameters taken from low-energy physics. This allows to relate
specific features of the confinement mechanisms with diffractive
electro-production processes and structure functions. The agreement
with phenomenological data is satisfactory.
\end{abstract}
\newpage
\setcounter{page}{1}
{\parindent0em \Large \bf Introduction}

In strong interaction physics the infra-red behavior of QCD is relevant
both in the low- and high-energy domain. In hadron spectroscopy this is
evident, but also in high-energy reactions at small momentum transfer
non-perturbative effects play a crucial role. Normally the
non-perturbative input in low-energy physics (e.g.~intra-quark
potentials) and high-energy physics (e.g.~structure functions) are
conceptually and computationally seen quite separately. In this
letter we calculate the high-energy diffractive scattering of
color-dipoles on a nucleon in a non-perturbative model of QCD, the
parameters of which are fixed only by low-energy input. Our results
are compared with other approaches and experimental data.

Since several years one of the authors (H.G.D.) and Yu.A.~Simonov have
proposed a model in which the non-perturbative features of QCD are
approximated by a Gaussian stochastic process which is characterized
by the gauge-invariant non-local gluon field-strength correlator
\cite{Dosch:1987}-\nocite{Dosch:1988}\cite{Simonov:1988}. This
correlator is characterized by a correlation length $a$ and its
value at zero separation, the gluon condensate $<g^2FF>$. This model
of the stochastic vacuum (MSV) yields linear confinement
\cite{Dosch:1987} in a very simple and natural way; a nice feature
of the model is that confinement only occurs in non-Abelian theories.

The MSV has also been applied to diffractive high-energy scattering
\cite{Dosch:1994}-\nocite{Rueter:1996}\cite{Rueter:1996III},
lepto-production of vector-mesons \cite{Dosch:1997} and
$\ga^*p$-interaction \cite{Dosch:1997II}. Here the natural ingredient
(see next section for more details) is the expectation value of two
Wegner-Wilson-loops with light-like sides, which correspond to the
world-lines of two color-neutral dipoles \cite{Nachtmann:1991}
\cite{Nachtmann:1996}\cite{Dosch:1994}. A characteristic feature of
the application of the MSV to diffractive high-energy scattering is
the fact that the same mechanism, which is responsible for confinement,
leads to a typical dependence of the cross-sections on the loop
extensions (dipole sizes): Even if the dipole size is large as
compared to the correlation length $a$ of the correlator, the total
cross-section still increases with the dipole size. Thus the model
implies string-string-interaction and can not lead to quark-additivity.
Nevertheless, the ratio of the total cross-sections off
nucleon-nucleon- and nucleon-meson-scattering agrees with experiment
without additional parameters, this is due to the different radii of
the nucleon and mesons, given by the electro-magnetic form-factor
\cite{Dosch:1994}. This dependence on the radii also leads to a typical
relation between the total cross-section and the logarithmic slope of
the elastic cross-section for different hadrons which is in agreement
with the experimental data \cite{Dosch:1994}.

In this letter we concentrate on the high-energy diffractive scattering
of a color-dipole of arbitrary size on a proton. This process is
phenomenologically very interesting because it can be used to describe
for example the diffractive electro-production of vector-mesons, where
due to the photon wave-function (in light-cone perturbation theory
\cite{Bjorken:1971}\cite{Lepage:1980}) dipoles of arbitrary size  occur
in the interaction. For the proton we use a diquark-picture, which works
in all our applications to high-energy scattering better than a
three-body-picture. Especially the necessary suppression of the
coupling of the odderon to the nucleon is achieved in that way
\cite{Rueter:1996}\cite{Rueter:1996III}. So, the electro-production
of vector-mesons can be build from dipole-dipole-scattering and is
illustrated in \fig{electroproduction}.
\epsfxsize8cm
\epsfysize4cm
\befi{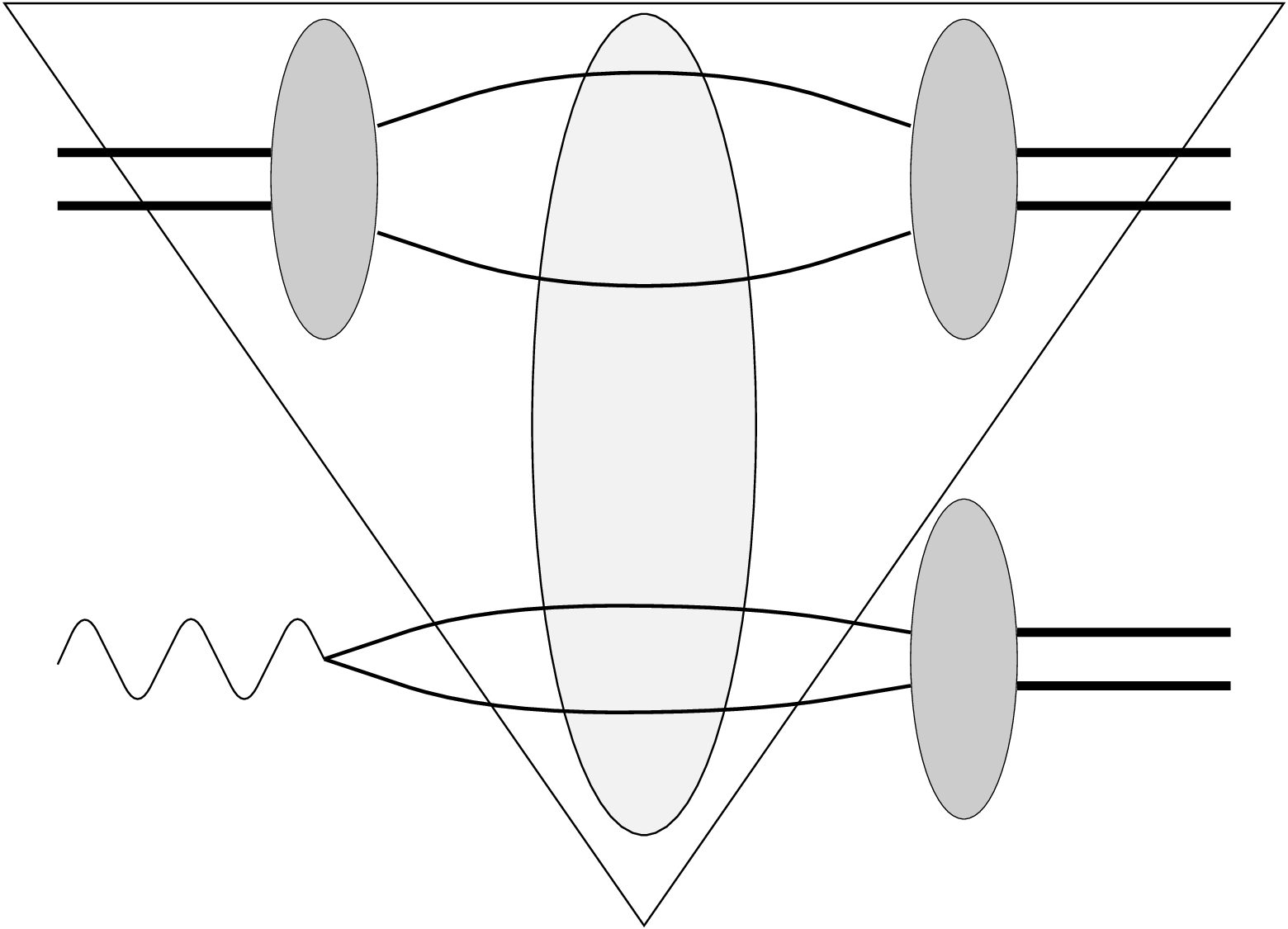}
\unitlength1cm
\begin{picture}(0,0)
\small
\put(-7.5,0.5){$Q^2=-q^2$}
\put(-1.5,0.5){$\rh,\,\om,\,\Ph\dots$}
\put(-4.5,2.3){$s$ large}
\put(-4.5,1.8){$t\le 1\GeV^2$}
\put(-7.3,3.6){proton}
\put(-1.6,3.6){proton}
\end{picture}
\lbcap{12cm}{Diffractive electro-production of vector-mesons can be
viewed as the diffractive scattering of color-dipoles ($q$-$\bar{q}$-pair
for the photon and $q$-diquark-pair for the proton) folded with
wave-functions. In this letter we calculate only the part inside the
triangular, that is the diffractive scattering of the proton on a dipole
of arbitrary size.}{electroproduction}\\
For large $Q^2$ the virtual photon (at least the longitudinal polarized
one) interacts like a small dipole. In the phenomenological interesting
regime dipole sizes between 0.1 fm and 1 fm are most interesting, where
the cross-over from perturbative to non-perturbative physics
occurs.\\[.5cm]
{\parindent0em \Large \bf Dipole-proton-scattering amplitude}

Nachtmann has developed a scheme which allows the separation of the hard
high-energy scale from the soft scale of momentum transfer
\cite{Nachtmann:1991}. He considered quark-(anti-)quark-scattering first
in an external gluon field in the eikonal approximation. In this way
each (anti-)quark picks up the eikonal phase
\beq
\cP e^{-ig\int_{C} \bA_\mu(x){\rm d}x^\mu}
\lbq{phase}
where the path $C$ is the classical (nearly) light-like path of the
(anti-)quark, $\bA_\mu$ is the matrix-valued color-potential and $\cP$
stands for path ordering. The quantum transition amplitudes are obtained
by functional integration over the color-field with the QCD-action as
exponential weight. In our treatment of high-energy scattering we want
to apply the MSV for this integration. We therefor consider not
quark-quark-scattering but rather scattering of color-neutral dipoles.
In that way we obtain from the eikonal factors (\eq{phase})
Wegner-Wilson-loops, i.e.~traces of line integrals around closed loops
whose light-like sides are the paths of the constituents of the dipole
(see \fig{2loops}). The scattering amplitude of two dipoles with
transversal extension $\vec{R}_1$ and $\vec{R}_2$ respectively and
impact parameter $\vec{b}$ is given by:
\beq
\tilde{J}(\vec{b},\vec{R}_1,\vec{R}_2)=-<\frac{1}{N_C}\Tr \left[ \bW
[\pa S_1]-{\bf 1}\right]\cdot\frac{1}{N_C}\Tr \left[ \bW [\pa S_2]-
{\bf 1}\right]>.
\lbq{Jredmeson}
\epsfxsize6cm
\befi{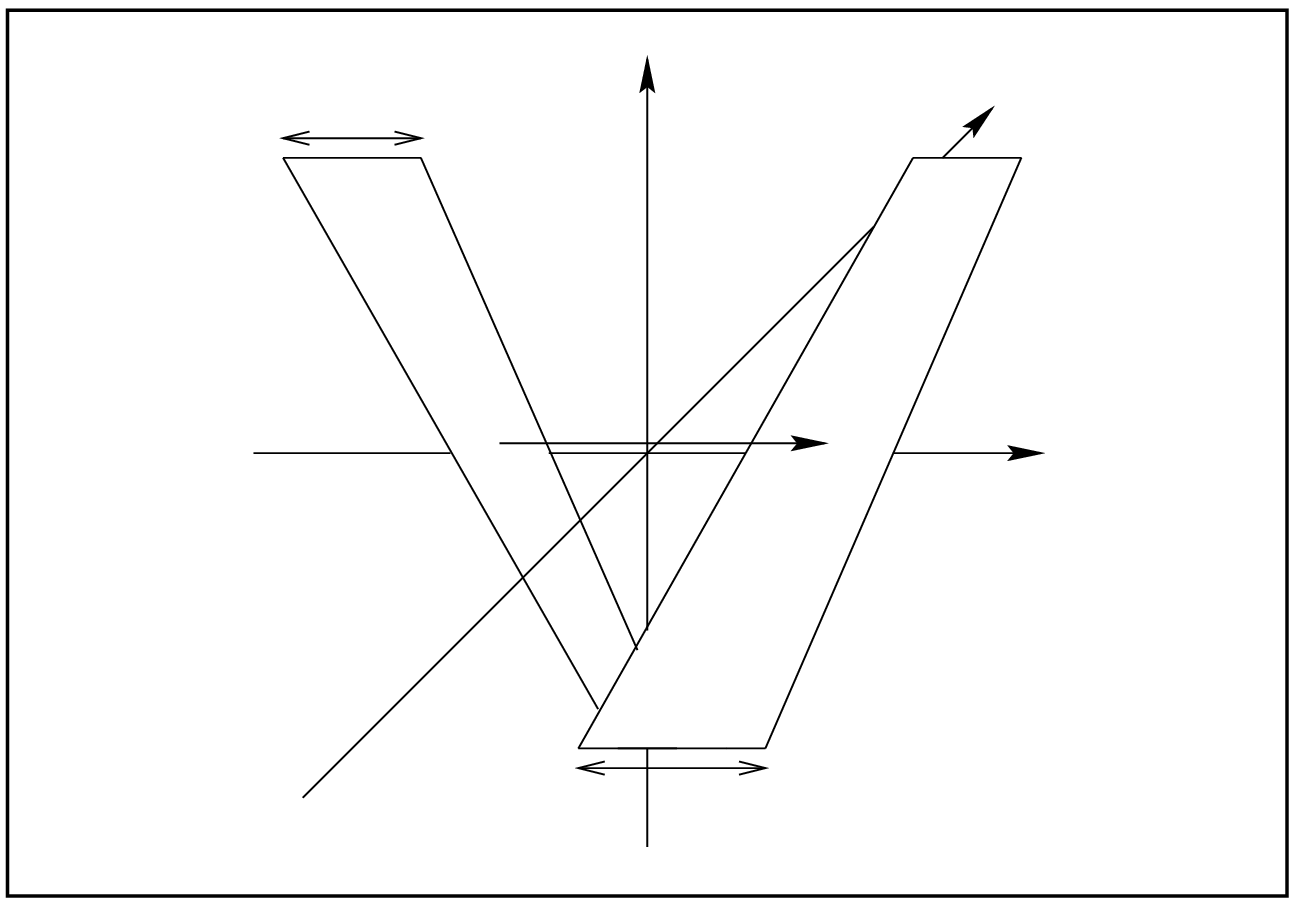}
\unitlength.6cm
\begin{picture}(0,0)
\put(-9.1,4.5){loop 1}
\put(-2,4.5){loop 2}
\put(-2.3,2.9){$\vec{x}$}
\put(-4.8,6.2){$x^0$}
\put(-2.2,6){$x^3$}
\put(-7.5,6.2){$\vec{R}_1$}
\put(-4.9,.4){$\vec{R}_2$}
\put(-5.5,3.7){$\vec{b}$}
\end{picture}
\lbcap{12cm}{The two W-loops with transversal extension $\vec{R}_1$ and
$\vec{R}_2$ and light-like sides. Loop 1 describes a color-dipole moving
in negative $x^3$-direction and loop 2 a color-dipole moving in positive
$x^3$-direction. The impact parameter $\vec{b}$ is chosen to be purely
transversal.}{2loops}
The brackets ($<\dots >$) denote the functional integration over the
gluon-field. In order to perform this integration we first transform
the line integrals over $\bA_\mu$ along the loops $\pa S_1,\,\pa S_2$
in surface integrals over the field-strengths and then expand the
exponentials. For the resulting functional integration over the
field-strengths we use the ansatz of the MSV \cite{Dosch:1987}
\cite{Dosch:1994}:
\beqa
&&<g^2\, F_{\mu\nu}^A(z,w)\,F_{\rh\si}^B(0,w)>\nn\\
&=&\frac{\de^{AB}}{N_C^2-1}\frac{<g^2FF>}{12}\Big\{ \ka \left(
g_{\mu\rh}g_{\nu\si}-g_{\mu\si}g_{\nu\rh} \right) D(z)\nn\\
&& +(1-\ka )\frac{1}{2} \left[ \pa_\mu \left( z_\rh g_{\nu\si}-z_\si
g_{\nu\rh} \right) + \pa_\nu \left( z_\si g_{\mu\rh}-z_\rh g_{\mu\si}
\right) \right] D_1(z) \Big\}.
\lbqa{Korrelator}
The correlation functions $D$ and $D_1$ are fitted to lattice results
of the non-local correlator \cite{DiGiacomo:1992}\cite{DiGiacomo:1996},
they fall off on a length scale given by the correlation length $a$.\\
In leading order of the expansion of the W-loops we obtain a real
scattering amplitude where only transversal coordinates enter. In
ref.~\cite{Dosch:1994} only the $D$-function of the non-local correlator
was taken into account. Including the $D_1$-function we find for
\eq{Jredmeson}:
\beqa
\tilde{J}&=&\frac{1}{8 N_C^2(N_C^2-1)144}\tilde{\ch}^2\nn\\
\tilde{\ch}&=&<g^2FF>\left( \ka \, \int_0^1{\rm d}w_1\, \int_0^1
{\rm d}w_2\, \left[ \vec{r}_{1q}\cdot\vec{r}_{2q}\, \cF_2[i\tilde{D}]
\left(w_1 \vec{r}_{1q}-w_2\vec{r}_{2q}\right)\right. \right.\nn\\
&&\hspace{5.1cm}+\vec{r}_{1\bar{q}}\cdot\vec{r}_{2\bar{q}}\,
\cF_2[i\tilde{D}]\left(w_1\vec{r}_{1\bar{q}}-w_2\vec{r}_{2\bar{q}}
\right)\nn\\
&&\hspace{5.1cm}-\vec{r}_{1q}\cdot\vec{r}_{2\bar{q}}\, \cF_2[i\tilde{D}]
\left(w_1\vec{r}_{1q}-w_2\vec{r}_{2\bar{q}}\right)\nn\\
&&\left.\hspace{5.1cm}-\vec{r}_{1\bar{q}}\cdot\vec{r}_{2q}\, \cF_2[i
\tilde{D}]\left(w_1\vec{r}_{1\bar{q}}-w_2\vec{r}_{2q}\right)\right]\nn\\
&&\hspace{1cm}+ (1-\ka ) \left[ \cF_2[i\tilde{D}_1 ']\left(\vec{r}_{1q}-
\vec{r}_{2q}\right)\right. +\cF_2[i\tilde{D}_1 ']\left(\vec{r}_{1\bar{q}}-
\vec{r}_{2\bar{q}}\right)\nn\\
&&\hspace{2.4cm}-\cF_2[i\tilde{D}_1 ']\left(\vec{r}_{1q}-
\vec{r}_{2\bar{q}}\right)-\left.\cF_2[i\tilde{D}_1 ']\left(
\vec{r}_{1\bar{q}}-\vec{r}_{2q}\right)\right]\bigg).
\lbqa{chi}
The vectors $\vec{r}$ are defined in \fig{mesonmesontrans} and $\cF_2$
denotes the two dimensional Fourier-transformed of the correlation
functions $D$ and $D_1$ in the transversal plane. Using the explicit
ansatz given in ref.~\cite{Dosch:1994} we have:
\beqa
\cF_2[i\tilde{D}(\vec{k})](\vec{x})&=& \int\, \frac{{\rm d^2}k}{(2\pi)^2}
i\tilde{D}(\vec{k})e^{i\vec{k}\cdot\vec{x}}\nn\\
&=&\frac{\pi}{2}\left[ 6 \frac{|\vec{x}|^2}{\la^2} K_2\left(
\frac{|\vec{x}|}{\la}\right) -\frac{|\vec{x}|^3}{\la^3} K_3
\left(\frac{|\vec{x}|}{\la}\right) \right] \la^2\nn\\
\cF_2[i\tilde{D}_1 '(\vec{k})](\vec{x})&=&\pi \frac{|\vec{x}|^3}
{\la^3}K_3 \left(\frac{|\vec{x}|}{\la}\right) \la^4\nn\\
\la&=&\frac{8}{3\pi}a.
\lbqa{2dimfourier}
Here $K_\nu$ denotes the modified Bessel-functions of second order.
One of the $w$-integrations in \eq{chi} can be performed analytically
\cite{Dosch:1994}.
\epsfxsize6cm
\befi{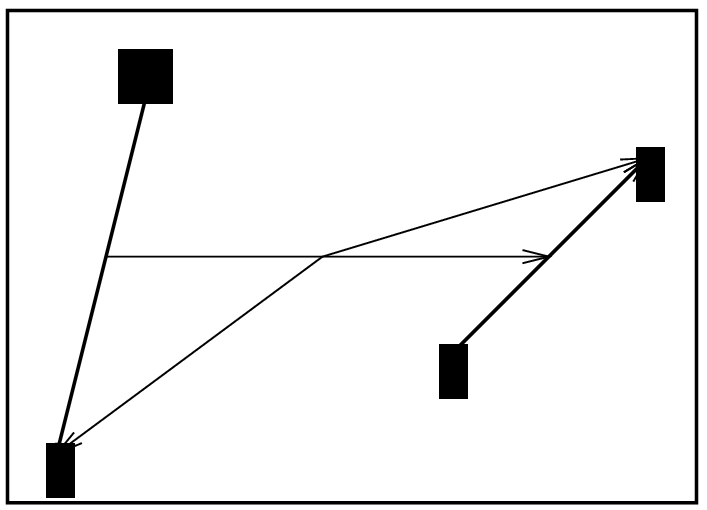}
\unitlength.8cm
\begin{picture}(0,0)
\put(-7.1,2.5){$\vec{R}_1$}
\put(-1.2,2.5){$\vec{R}_2$}
\put(-3.2,2.1){$\vec{b}$}
\put(-2.8,3.5){$\vec{r}_{2\bar{q}}$}
\put(-5.1,1.5){$\vec{r}_{1q}$}
\end{picture}
\lbcap{12cm}{Dipole-dipole-scattering in the transversal plane:
$\vec{R}_i$ points from the quark of dipole $i$ to the anti-quark;
$\vec{b}$ is the impact parameter and $\vec{r}_{iq(\bar{q})}$ points
from the origin to the (anti-)quark of dipole $i$.}{mesonmesontrans}\\
We also include the perturbative interaction which dominates the short
distance behavior of the correlation function $D_1$. We use the
perturbative expression for the non-local correlator in first order with
a infra-red cutoff in agreement with the lattice results
\cite{DiGiacomo:1996}:
\beqa
&&<g^2\, F_{\mu\nu}^A(z)\,F_{\rh\si}^B(0)>_{\rm per}\\
&=&\de^{AB}g^2\Ikdv e^{-ikz}\left( -g_{\nu\si}k_\mu k_\rh +
g_{\nu\rh}k_\mu k_\si-g_{\mu\rh}k_\nu k_\si+g_{\mu\si}k_\nu k_\rh \right)
\tilde{D}_{\rm per}'(k)\nn
\lbqa{korper}
with
\[
\tilde{D}_{\rm per}'(k)=\frac{i}{k^2-\frac{1}{c^2}},\;c=0.45\fm.
\]
For the coupling constant we use here a frozen value of
$\al_s\approx 0.5$, which is prescribed by our model for consistency
reasons \cite{Rueter:1995}-\nocite{Dosch:1995}\cite{Rueter:1996II}.
The perturbative contribution results in an additional term to
$\tilde{\ch}$ (\eq{chi}):
\beqa
\tilde{\ch}_{\rm per}&=&-\frac{12i}{4}\int_{S_{1}}\int_{S_{2}} <g^2
\, F_{\mu\nu}^A(x,w)\,F_{\rh\si}^A(y,w)>_{\rm per}\, {\rm d}
\si^{\mu\nu}(x)\, {\rm d}\si^{\rh\si}(y)\nn\\
&=&12(N_C^2-1)g^2\left[ \De\left(\vec{r}_{1q}-\vec{r}_{2q}\right)
\right. +\De\left(\vec{r}_{1\bar{q}}-\vec{r}_{2\bar{q}}\right)-\nn\\
&&\hspace{2.7cm}\De\left(\vec{r}_{1q}-\vec{r}_{2\bar{q}}\right)-\left.
\De\left(\vec{r}_{1\bar{q}}-\vec{r}_{2q}\right)\right]
\lbqa{chiper}
with
\beq
\De(\vec{x})=\cF_2\left[\frac{1}{\vec{k}^2+\frac{1}{c^2}}\right]
(\vec{x})=\frac{1}{2\pi}K_0\left[\frac{|\vec{x}|}{c}\right].
\lbq{2dimfourierper}
Because the function $\tilde{\ch}$ enters quadratically in the scattering
amplitude (see \eq{chi}) we have in addition to the purely perturbative
and non-perturbative contribution also an interference term.\\
In order to come from dipole-dipole- to dipole-hadron-scattering we smear
the scattering amplitude (\eq{Jredmeson}) with a transversal proton
wave-function with extension parameter $S_p$ and keep one extension
fixed to the dipole size $R_{\rm D}$:
\beq
\hat{J}(\vec{b},R_{\rm D},S_p)=\int\,\frac{{\rm d}\varphi_{\rm D}}{2\pi}
\int\,{\rm d^2}R_2\,\tilde{J}(\vec{b},\vec{R}_{\rm D},\vec{R}_2)
|\Ps(\vec{R}_2,S_p)|^2.
\lbq{T3B}
The scattering amplitude at center of mass energy $s$ and momentum
transfer \mbox{$t=-\vec{q}\,^2$} is given by:
\beq
T(s,t) = 2is \int \,{\rm d^2}b\,e^{-i\vec{q}\,\vec{b}}\,\hat{J}
(\vec{b},R_{\rm D},S_p).
\lbq{T}
For the proton (working in the diquark-picture \cite{Rueter:1996}
\cite{Rueter:1996III}) we use as wave-function:
\beq
|\Ps(\vec{R}_2,S_p)|^2= \frac{1}{2\pi}\frac{1}{S_p^2}e^{-\frac{
|\vec{R}_2|^2}{2S_p^2}}.
\lbq{Gausswfmeson}
For the total cross-section follows:
\beq
\si^{\rm tot}=\frac{1}{s}{\rm Im}T(s,0)=2\int \,{\rm d^2}b\,{\rm Re}
[\hat{J}(\vec{b},R_{\rm D},S_p)],
\lbq{sigtotundslope}
which in our model is independent of the center of mass energy $s$.

{\parindent0em \Large \bf Numerical results and comparison with other
approaches and experimental data}

Nikolaev and Zakharov developed a picture of dipole-nucleon scattering
\cite{Nikolaev:1991}-\nocite{Nikolaev:1992}\nocite{Nikolaev:1994}
\cite{Nikolaev:1994II} which is in some respect very similar to the
one discussed above. But whereas our main emphasis lies on the
calculation of the scattering amplitude $\hat{J}(\vec{b},R_{\rm D},S_p)$
(\eq{T3B}) they are more interested in the consequences of the
dipole-scattering picture especially in the energy dependence due to
perturbative QCD. In a recent analysis Nemchik et al.~extracted from
electro-production data the dipole-proton cross-section
$\si^{\rm tot}(s,R_{{\rm D}1})$ \cite{Nemchik:1996}\cite{Nemchik:1996II}.
They used the fact that due to the overlap of the wave-functions of
the photon with virtuality $Q^2$ and the produced vector-meson of mass
$M_V$ the dipole-nucleon cross-section is manly probed at the so called
scanning radius \cite{Nemchik:1994}
\[
r_s\approx \frac{6}{\sqrt{Q^2+M_V^2}}.
\]
The results of there analysis of different experiments are shown in
\fig{ohnefitten}. There are two distinct domains of energy of the
$\ga^*-p$-system: The fixed target experiments of the EMC, NMC, E687
and FNAL groups are at the center of mass energy about $W\approx 10-15$
GeV, whereas the HERA experiments are at $W\approx 70-150$ GeV. The
high-energy results are indeed systematically higher than the fixed
target results but here we do not discuss this point which is off
course of great interest for itself.
\epsfxsize10cm
\epsfysize8cm
\befi{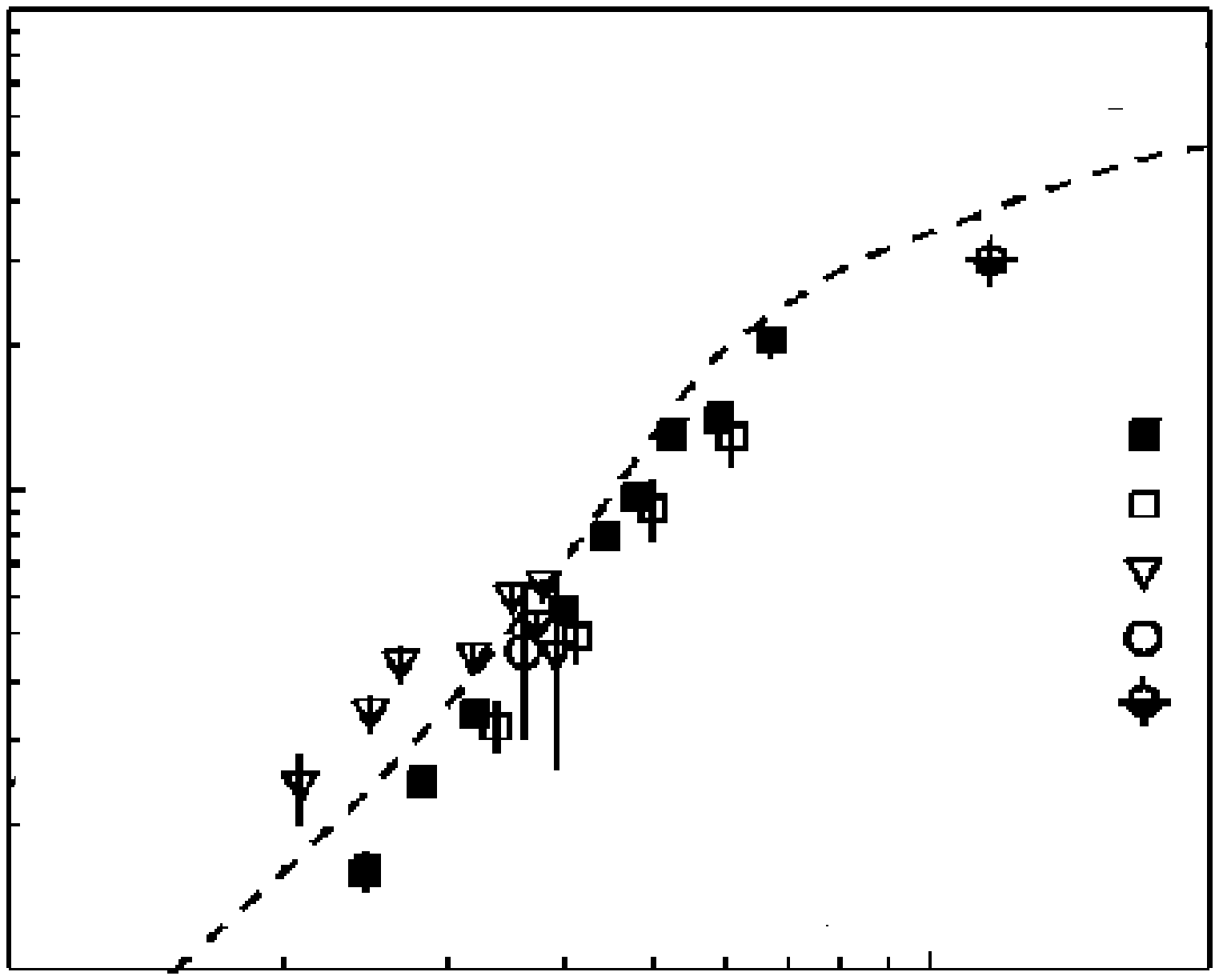}
\begin{picture}(0,0)
\put(-280,5){\epsfysize7.65cm\epsfxsize9.63cm\epsffile{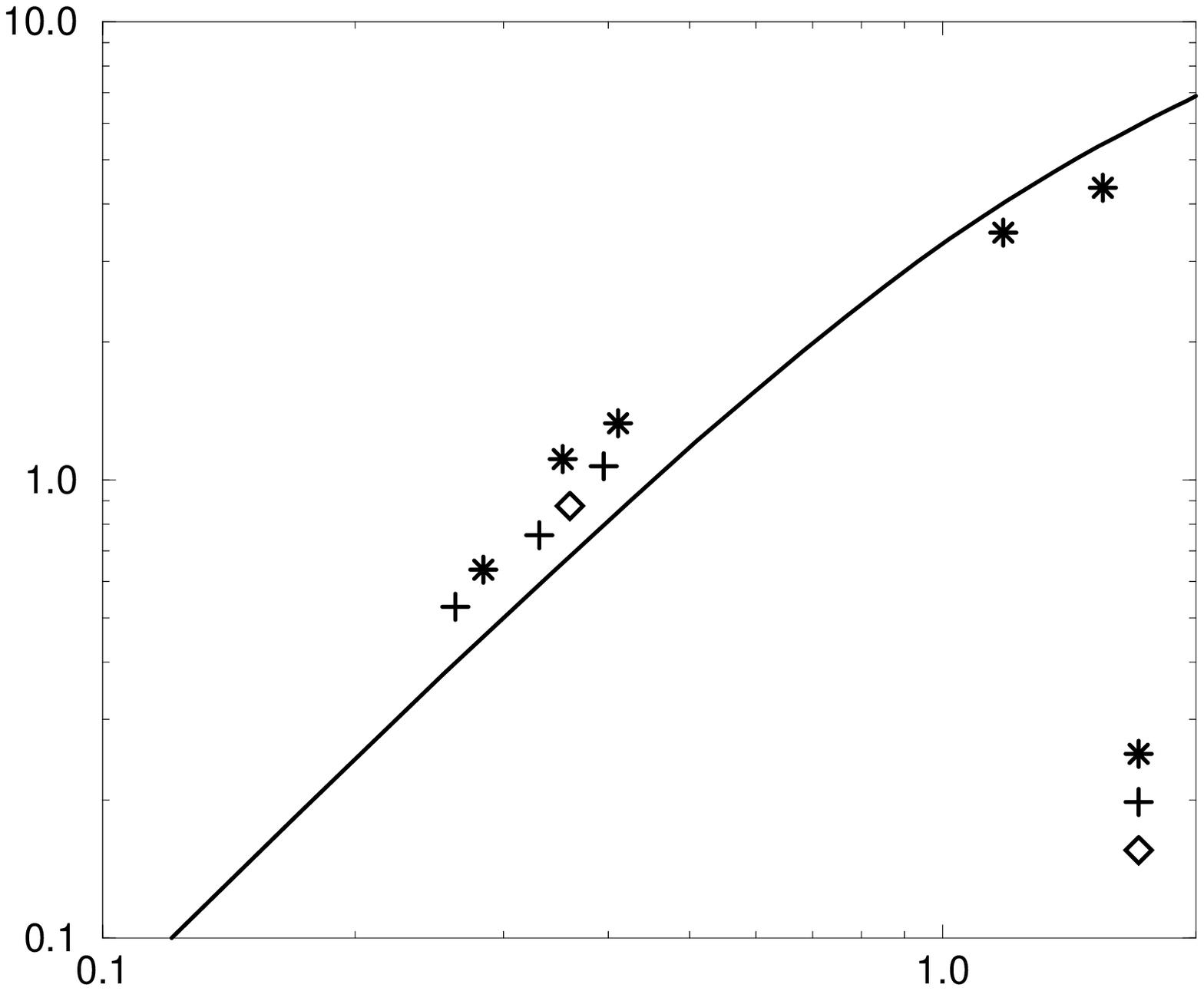}}
\put(-1.5,0){$R_{{\rm D}1}[\fm]$}
\put(-330,205){$\si^{\rm tot}[\fm^2]$}
\put(-95,125){$\rh^0$-NMC}
\put(-95,110){$\Ph^0$-NMC}
\put(-95,95){$J/\Psi$-EMC}
\put(-95,80){$J/\Ps$-E687}
\put(-95,65){$\Ph^0$-FNAL}
\put(-95,50){$\Ph^0,\,\rh^0$-ZEUS}
\put(-95,35){$\rh^0$-H1}
\put(-95,20){$J/\Ps$-HERA}
\end{picture}
\begin{minipage}{12cm}
\caption{\small Comparison of our result for the total cross-section for
dipole (extension $R_{{\rm D}1}$) proton scattering with values extracted
from cross-sections of lepto-production of vector-mesons by the method
of Nemchik et al.~\protect\cite{Nemchik:1996}\protect
\cite{Nemchik:1996II}. The solid line is our result without any fitting
of the parameters to high-energy data. The dashed line is the ansatz of
Nemchik et al.~for the total cross-section for the fixed target
experiments.}
\labelm{ohnefitten}
\end{minipage}
\end{figure}\\
The solid line in \fig{ohnefitten} is our prediction for the dipole-proton
cross-section where only low-energy input has been used: the correlator
(\eq{Korrelator}) taken from the lattice calculations (compatible with
the string tension of heavy quarkonia) and the electro-magnetic proton
radius (\mbox{$R^{\rm em}=0.862$ fm} \cite{Simon:1980}). We see that the
agreement with the extracted phenomenological points is satisfactory and
encourages the approach to relate high- and low-energy data.\\
Of special interest is the agreement of our calculation at values of
$R_{{\rm D}1}$ larger than the correlation length $a\approx 0.3$ fm.
If quark-additivity would hold the dipole cross-section should level
off around that value of $R_{{\rm D}1}$. The continued increase shows
that the most specific feature of non-perturbative QCD, namely confinement,
leaves its traces also in high-energy scattering.

{\parindent0em \Large \bf Acknowledgments}

The authors want to thank A.~Hebecker, O.~Nachtmann and H.J.~Pirner for
discussions. M.R.~thanks the Deutsche Forschungsgemeinschaft for financial
support.

\end{document}